\documentclass[copyright,creativecommons]{eptcs}

\usepackage{graphicx}

\usepackage{pdfpages}

\usepackage{float}

\usepackage{amssymb}
\usepackage{amstext}
\usepackage{amsmath}

\providecommand{\event}{ThEdu'18 Post-Proceedings} 
\usepackage{underscore}           

\title{Natural Deduction Assistant (NaDeA)}
\author{J{\o}rgen Villadsen \email{} \and Andreas Halkj{\ae}r From \email{} \and Anders Schlichtkrull \email{}
\institute{DTU Compute - Department of Applied Mathematics and Computer Science,\\[1ex]
Technical University of Denmark, Richard Petersens Plads, Building 324, DK-2800 Kongens Lyngby, Denmark}
}
\def\titlerunning{Natural Deduction Assistant (NaDeA)}
\def\authorrunning{J. Villadsen, A. H. From \&\ A. Schlichtkrull}

\usepackage{tabulary}

\begin{document}

\maketitle 

\begin{abstract}
\noindent
We present the Natural Deduction Assistant (NaDeA) and discuss its advantages and disadvantages as a tool for teaching logic. NaDeA is available online and is based on a formalization of natural deduction in the Isabelle proof assistant.
We first provide concise formulations of the main formalization results.
We then elaborate on the prerequisites for NaDeA, in particular we describe a formalization in Isabelle of ``Hilbert's Axioms'' that we use as a starting point in our bachelor course on mathematical logic.
We discuss a recent evaluation of NaDeA and also give an overview of the exercises in NaDeA.
\end{abstract}

\section{Introduction}

The Natural Deduction Assistant (NaDeA) \cite{DBLP:journals/corr/abs-1803-01473,IfCoLog} runs in a standard browser:
\begin{center}
\url{https://nadea.compute.dtu.dk/}
\end{center}

\noindent
NaDeA is written in the TypeScript programming language and is open source (MIT License):
\begin{center}
\url{https://github.com/logic-tools/nadea}
\end{center}

\noindent
NaDeA is based on a formalization of natural deduction in the Isabelle proof assistant and the overall aim of the present paper is to discuss its advantages and disadvantages as a tool for teaching logic.

Our formalization in the Isabelle proof assistant \cite{nipkow+02} of the syntax, semantics and the inductive definition of the natural deduction proof system extends work by Berghofer \cite{berghofer} but with a much more detailed soundness proof that can be examined and tested by the students. The corresponding completeness proof is also available but it is of course quite demanding for a student to understand. NaDeA can be used with or without installing Isabelle and it is not necessary that the students have any knowledge about proof assistants \cite{geuvers}.

The present paper extends our previous publications about NaDeA \cite{DBLP:journals/corr/abs-1803-01473,IfCoLog}.
In section 2 we provide more concise formulations of the main formalization results and in sections 3 and 4 we describe a sample proof in NaDeA and list selected features for students.
In section 5 we elaborate on the prerequisites for NaDeA and we discuss a recent evaluation of NaDeA in section 6.
In section 7 we give an overview of the exercises in NaDeA and finally we conclude in section 8.

\newpage

\section{Main Formalization Results}

\begin{figure}
\begin{center}
\includegraphics[trim=10mm 20mm 75mm 5mm,clip,width=\textwidth,height=\textheight,keepaspectratio]{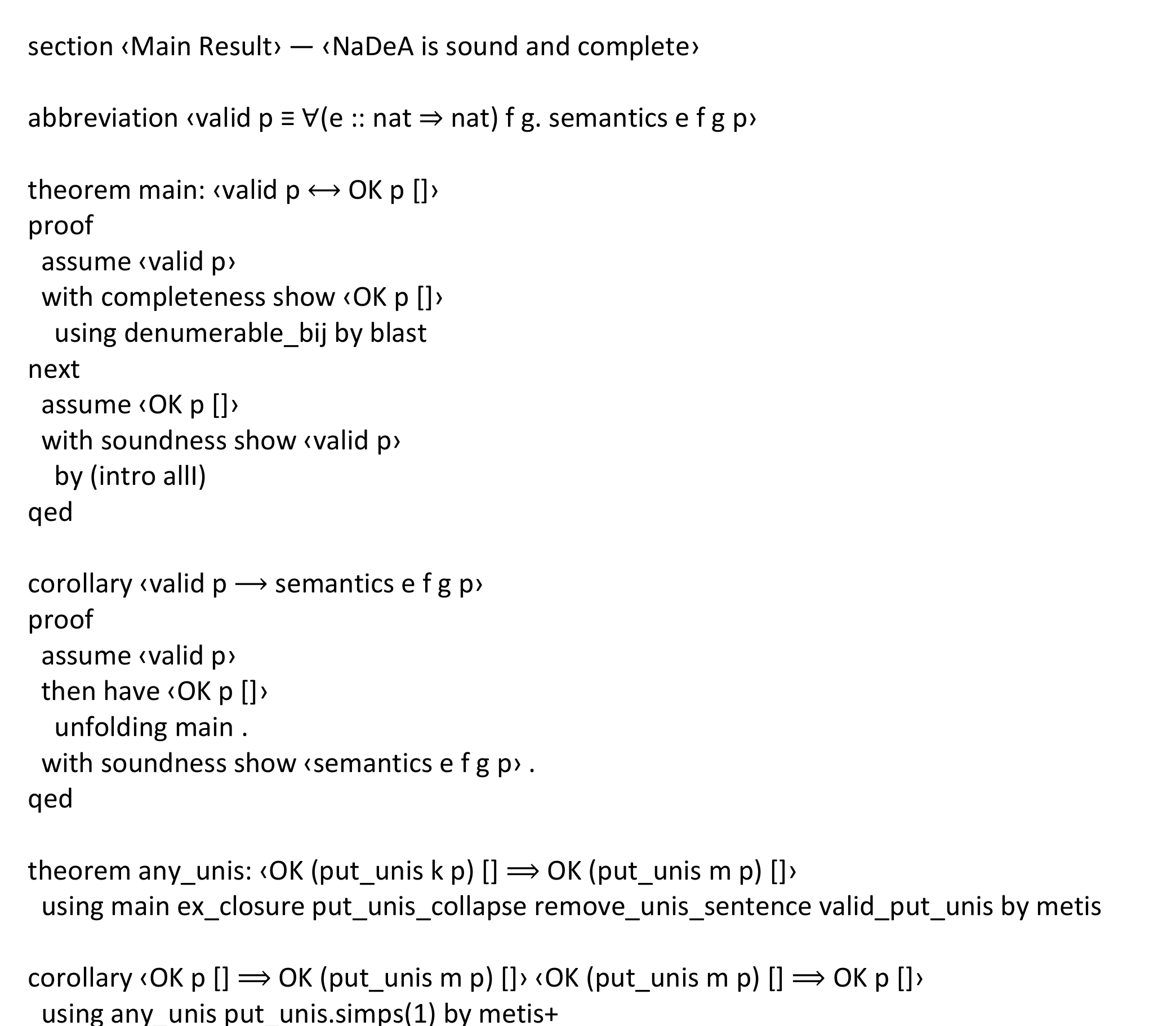}
\end{center}
\caption{Main Formalization Results \label{Main}}
\end{figure}

NaDeA is based on a formalization of natural deduction in the Isabelle proof assistant.
Figure \ref{Main} shows the main results which are more concise formulations of the formalization results discussed in our previous publications \cite{DBLP:journals/corr/abs-1803-01473,IfCoLog}.

We define the validity of a formula as the formula evaluating to true in all variable denotations (\textsf{e}), function denotations (\textsf{f}) and predicate denotations (\textsf{g}) with the natural numbers as universe. This is different from the usual notion of validity which considers all universes --- not only that of the natural numbers. We can, however, prove in Isabelle that our notion of validity implies the truth of a formula in any variable denotation, function denotation and predicate denotation --- and thus the formula must indeed be valid with respect to the usual notion of validity.

We prove that the valid formulas are exactly the same as those which can be proved using the provability predicate \textsf{OK}, hence the natural deduction proof system is sound and complete (theorem \textsf{main} in Figure \ref{Main}).
We prefer the word \textsf{OK} to longer words like ``provable'' or symbols like $\vdash$ mainly because it is easier to pronounce in class. The complete Isabelle theory file with proofs of soundness and completeness is checked by Isabelle/HOL in a few seconds. And from this theory loaded into Isabelle's front-end a mouse-click, for instance on \textsf{OK}, leads to the respective definition --- an invitation to investigate the underlying mechanized mathematics deeper and deeper. The more than 4000 lines are available here:
\begin{center}
\url{https://nadea.compute.dtu.dk/Natural_Deduction_Assistant.thy}
\end{center}

\noindent
The theorem \textsf{any-unis} is a result of having completeness for open formulas and states that given a proof with some number \textsf{k} of outer universal quantifiers, a proof with \textsf{m} quantifiers, either fewer or more, can be derived.
The theorem \textsf{any-unis} follows from \textsf{main} as well as the lemmas \textsf{ex\_closure}, \textsf{put\_unis\_collapse}, \textsf{remove\_unis\_sentence} and \textsf{valid\_put\_unis}.
When a formula has been proved in NaDeA a small Isabelle theory file is generated that verifies the validity of the original formula as well as the validity of all versions of it with some number of outer universal quantifiers omitted. The theorem \textsf{any-unis} is used by this generated Isabelle theory file.

\section{A Sample Proof}

We consider the following formula and its online proof:
\[
\exists x. A(x) \rightarrow (\forall x. A(x))
\]
The proof in NaDeA is obtained by clicking \textsf{Cancel help}, \textsf{Load}, \textsf{Test 9} and \textsf{Load shown proof}.
The formula is the so-called drinker paradox:
\begin{quote}
There is someone in the pub such that, if that person is drinking, then everyone in the pub is drinking.
\end{quote}
It was popularized by Raymond Smullyan:
\begin{center}
\url{https://en.wikipedia.org/wiki/Drinker_paradox}
\end{center}

\begin{figure}
\begin{center}
\includegraphics[trim=0mm 390mm 0mm 0mm,clip,width=\textwidth,height=\textheight,keepaspectratio]{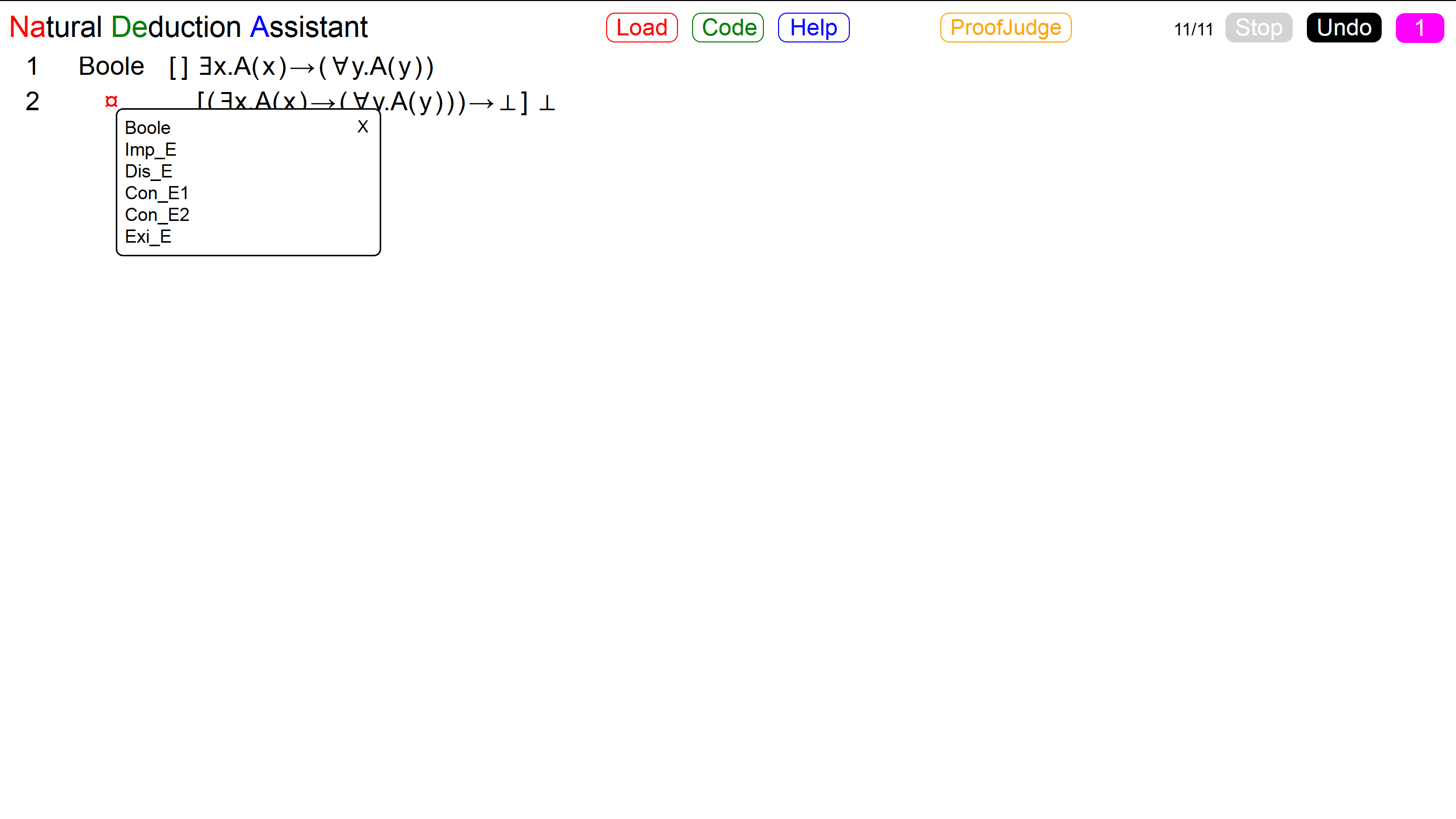}
\end{center}
\caption{A Sample Proof -- Start \label{N7}}
\end{figure}

\noindent
Figure \ref{N7} shows the start of the proof --- more or less --- where a natural deduction rule is to be chosen in the proof step 2.
This state can be obtained from the previous one by clicking \textsf{Undo} repeatedly -- one can always click on \textsf{Undo} to go back all the way to the very first proof step 1.

\begin{figure}
\begin{center}
\includegraphics[trim=0mm 145mm 0mm 0mm,clip,width=\textwidth,height=\textheight,keepaspectratio]{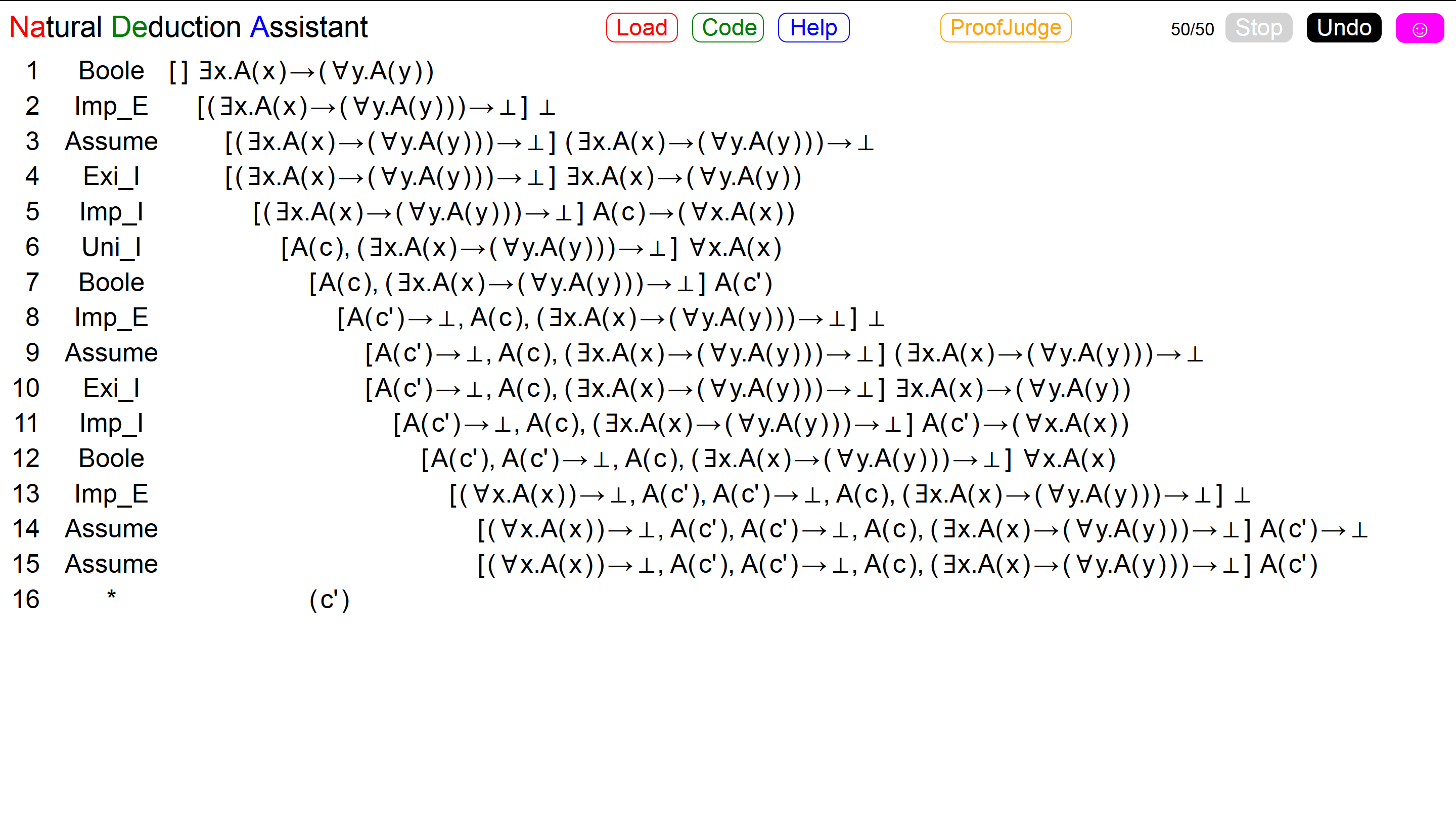}
\end{center}
\caption{A Sample Proof -- Finish \label{N8}}
\end{figure}

Figure \ref{N8} shows the finished proof which can be reached again by first clicking \textsf{Stop} and then \textsf{Undo} repeatedly.

A formula \textsf{[\,] p} in line 1 corresponds to the expression \textsf{OK p [\,]} in the previous section and the names of the natural deduction rules are the same as used in the formalization in Isabelle.

\section{Selected Features for Students}

\begin{figure}
\begin{center}
\includegraphics[width=\textwidth,height=\textheight,keepaspectratio]{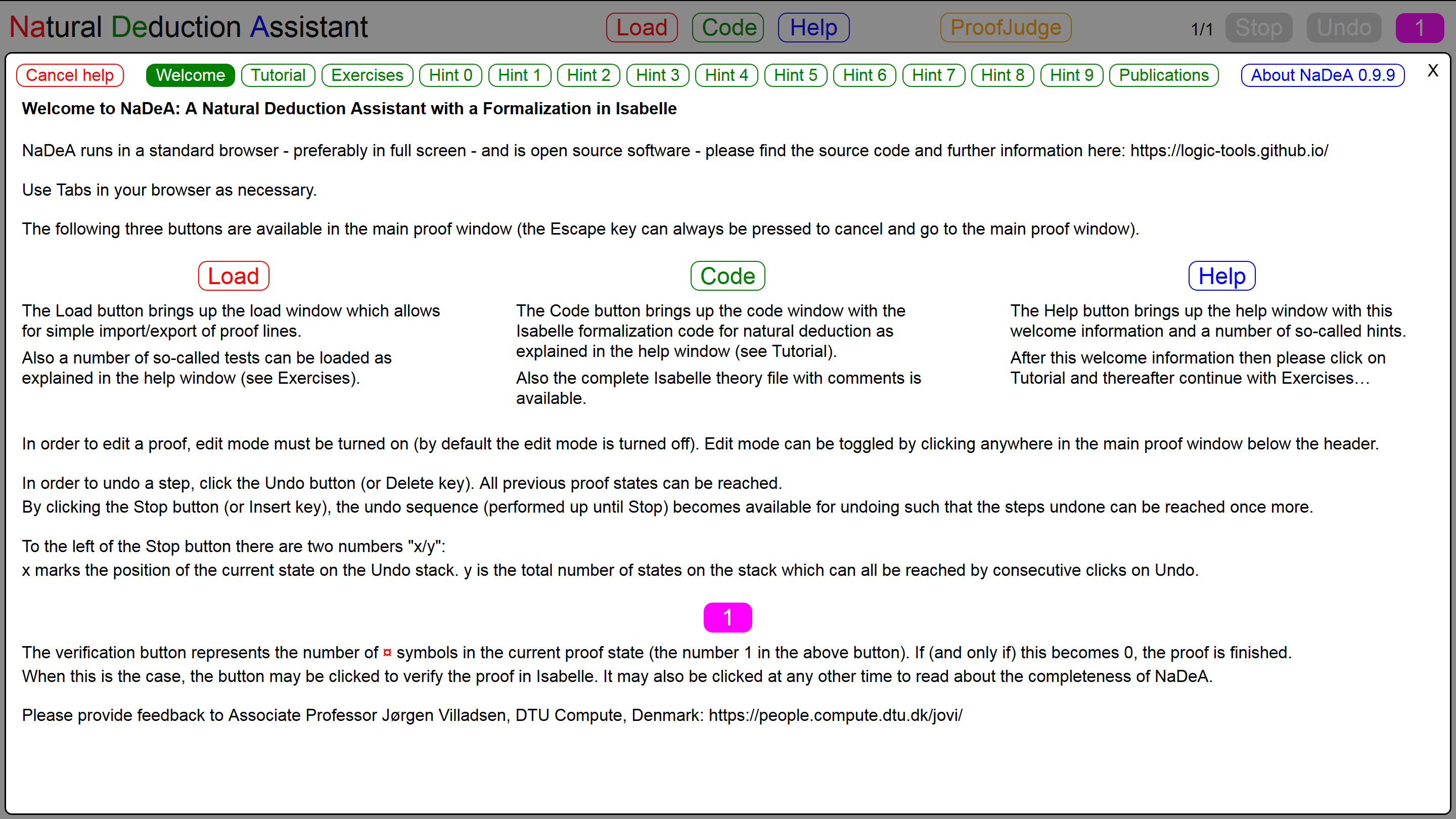}
\end{center}
\caption{Welcome \label{N1}}
\end{figure}

\begin{figure}
\begin{center}
\includegraphics[width=\textwidth,height=\textheight,keepaspectratio]{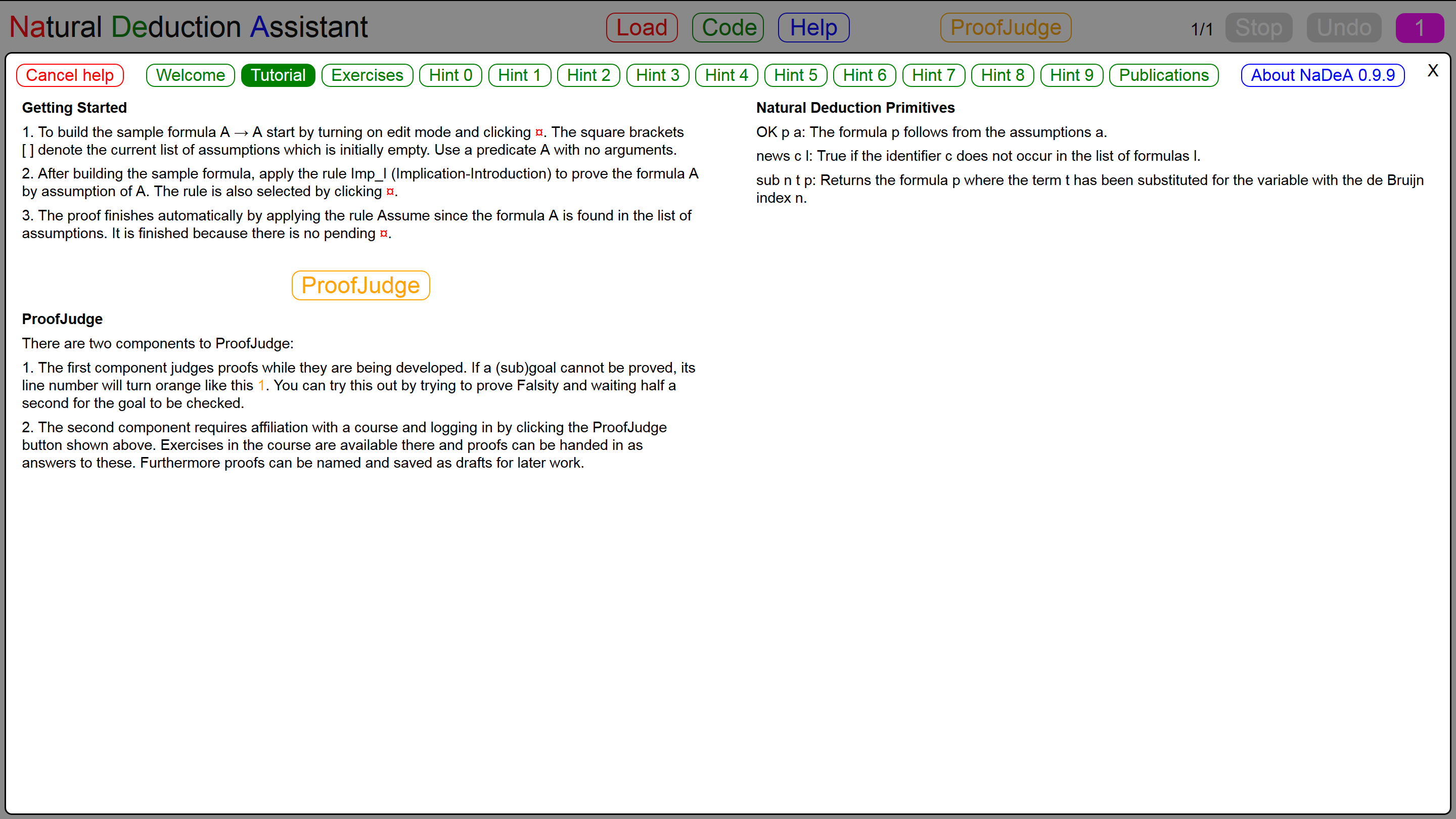}
\end{center}
\caption{Tutorial \label{N2}}
\end{figure}

\begin{figure}
\begin{center}
\includegraphics[width=\textwidth,height=\textheight,keepaspectratio]{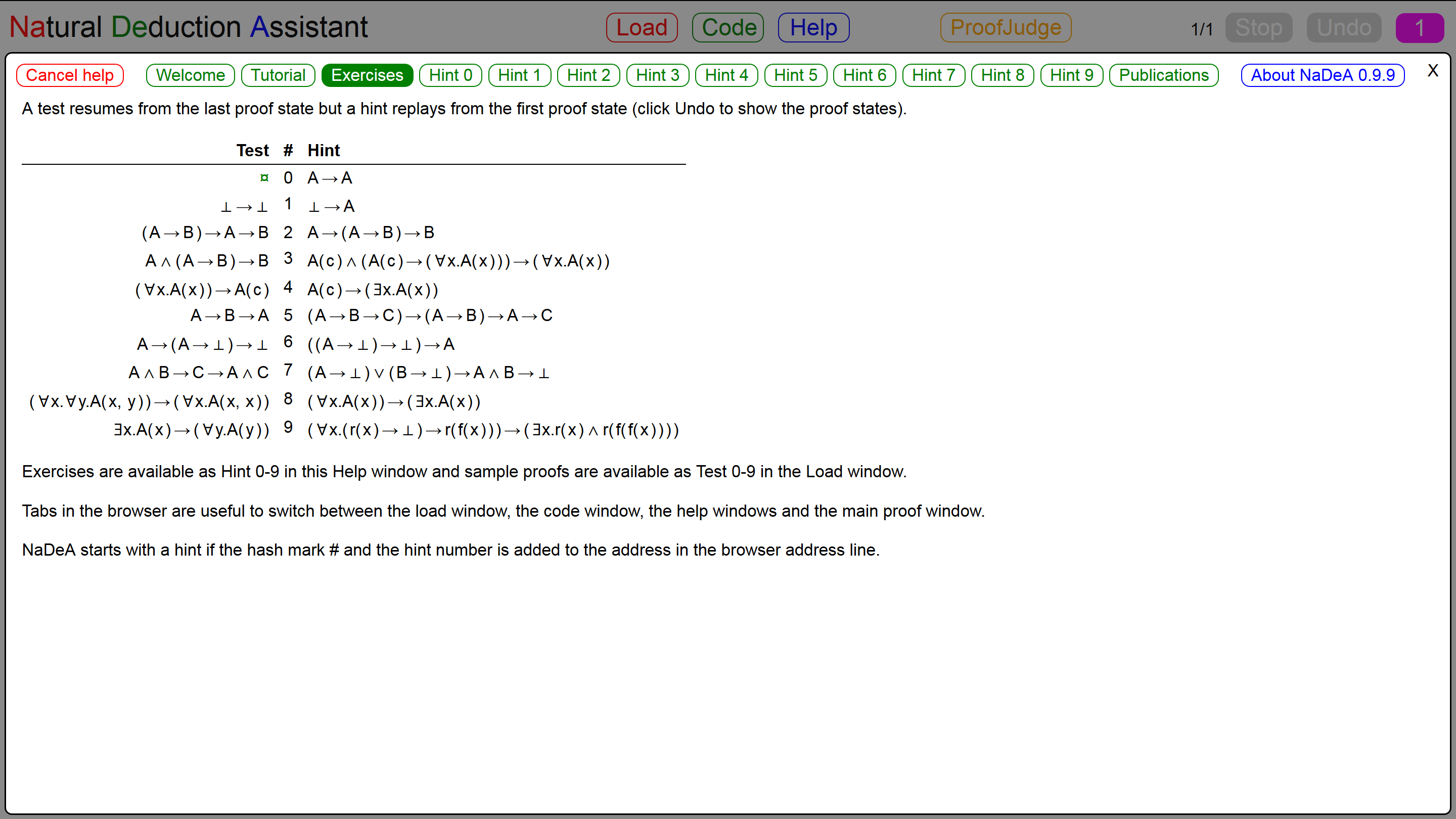}
\end{center}
\caption{Exercises \label{N3}}
\end{figure}

\begin{figure}
\begin{center}
\includegraphics[width=\textwidth,height=\textheight,keepaspectratio]{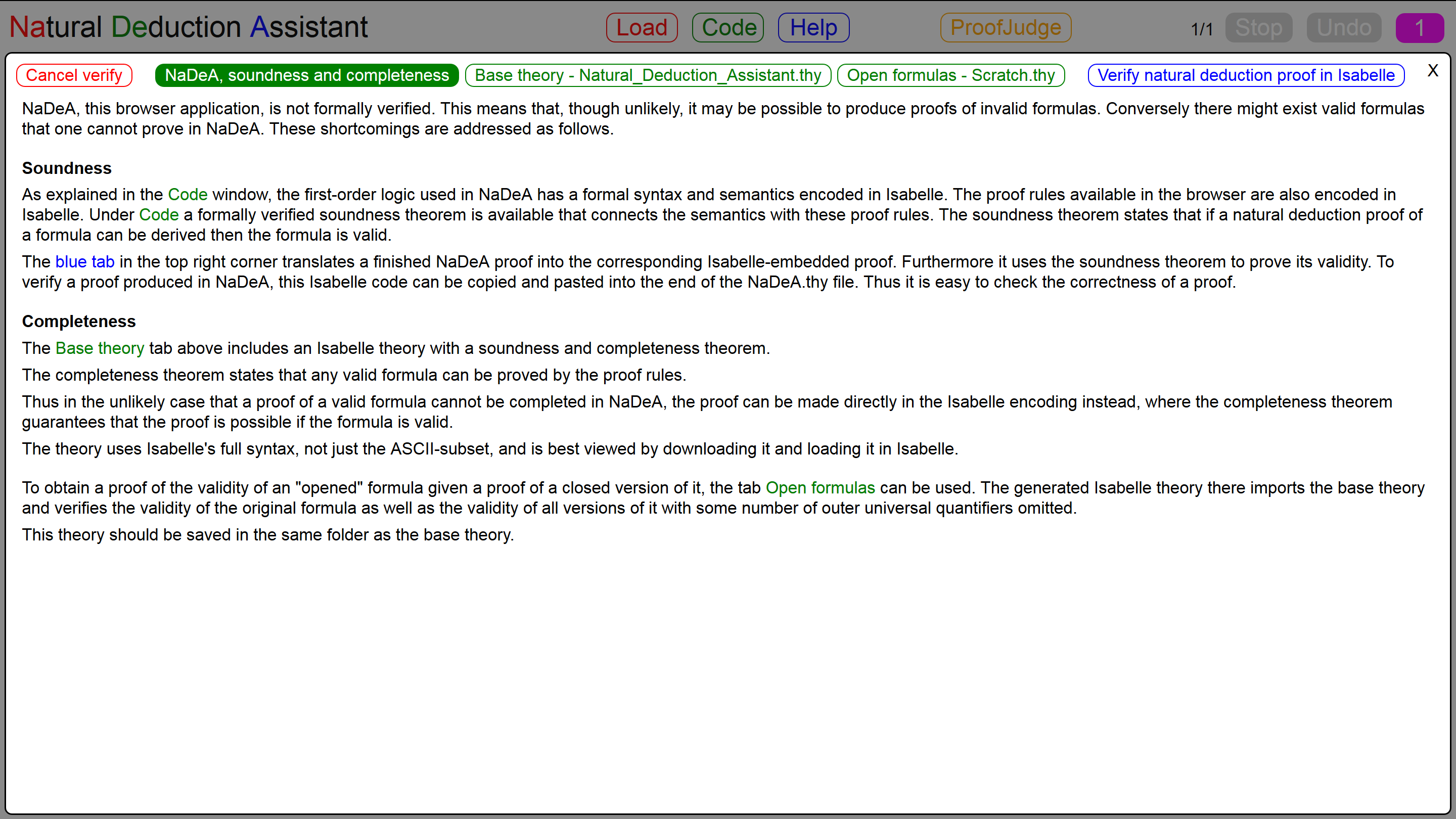}
\end{center}
\caption{NaDeA, soundness and completeness \label{N4}}
\end{figure}

\begin{figure}
\begin{center}
\includegraphics[width=\textwidth,height=\textheight,keepaspectratio]{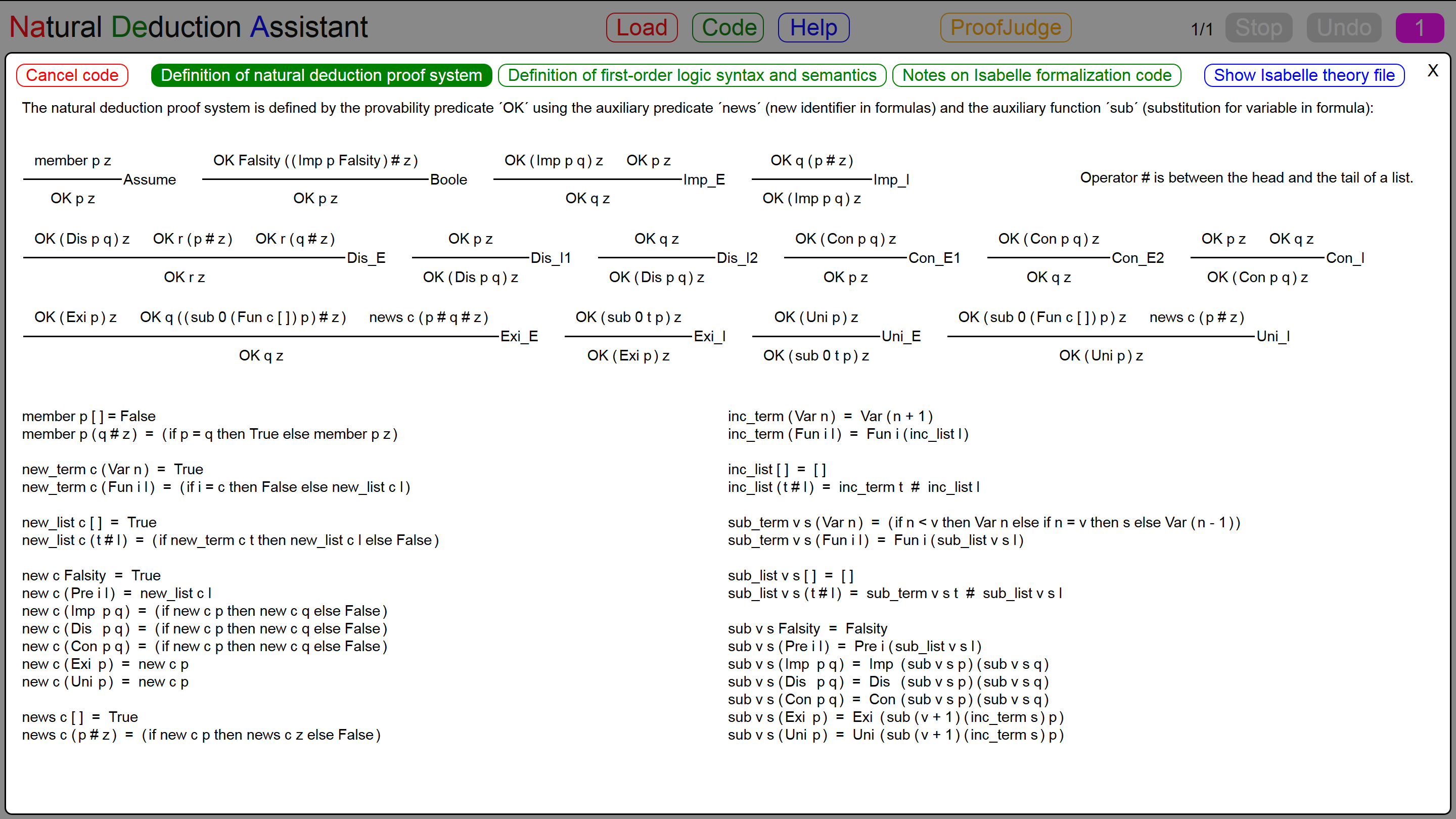}
\end{center}
\caption{Definition of natural deduction proof system \label{N5}}
\end{figure}

\begin{figure}
\begin{center}
\includegraphics[width=\textwidth,height=\textheight,keepaspectratio]{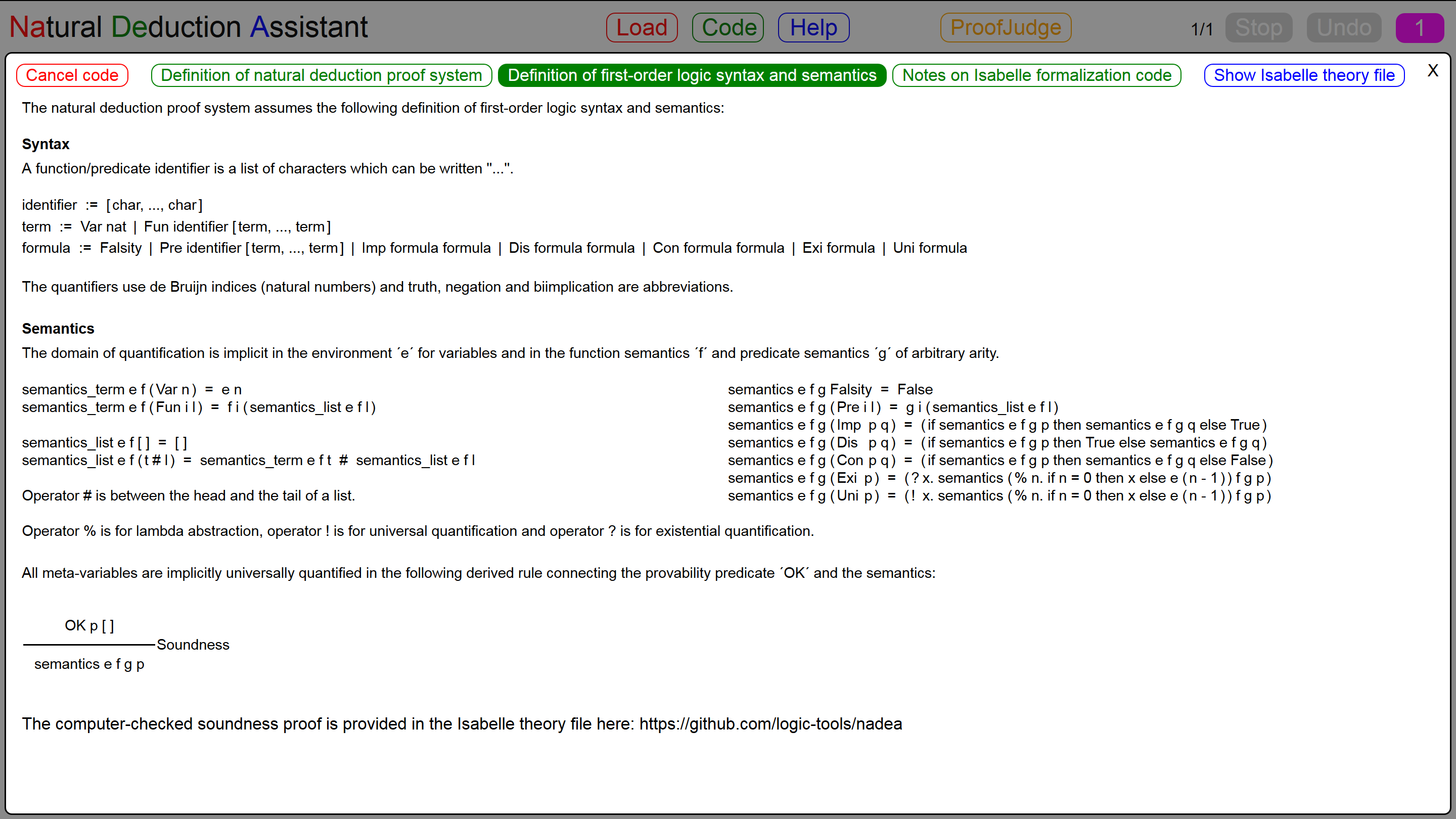}
\end{center}
\caption{Definition of first-order logic syntax and semantics \label{N6}}
\end{figure}

We briefly describe a number of NaDeA features for students:
\begin{itemize}
\item
Figure \ref{N1} shows the Welcome window. The Help button brings up the help window with this welcome information and a number of so-called hints.
\item
Figure \ref{N2} shows the Tutorial window. It contains a getting started guide as well as a list of the natural deduction primitives.
\item
Figure \ref{N3} shows the Exercises window. Solutions to all exercises are provided and can be revealed step-by-step with hints.
\item
Figure \ref{N4} shows the special NaDeA, soundness and completeness, window. The so-called verification button allows the user to verify any finished proof in Isabelle.
\item
Figure \ref{N5} shows the major Isabelle Code window -- entitled: Definition of natural deduction proof system -- with the formalization in Isabelle.
\item
Figure \ref{N6} shows the minor Isabelle Code window -- entitled: Definition of first-order logic syntax and semantics -- with the formalization in Isabelle.
\end{itemize}
There are several other NaDeA features for students -- for example, the ProofJudge system in NaDeA can manage student assignments in courses with teaching assistants.

NaDeA uses the automation of our verified declarative prover tool \cite{jensen2018} to give students feedback on the provability of their goals and subgoals.
Overall both that prover and NaDeA are related to the IsaFoL  project which unites researchers in formalizing logic in Isabelle:
\begin{center}
\url{https://bitbucket.org/isafol}
\end{center}

Among the formalizations in the project are SAT-solving, first-order resolution, a paraconsistent logic, sequent calculi and more.

\section{Student Prerequisites for NaDeA}

We use NaDeA in a bachelor course with the following prerequisites:
\begin{itemize}
\item Discrete mathematics (basic set theory)
\item Algorithms and data structures (searching and sorting)
\item Programming in a functional programming language 
\end{itemize}
We use 5 lectures (each 45 minutes) plus exercise sessions with teaching assistants:
\begin{enumerate}
\item Warm-up --- Truth Tables \&\ Isabelle Introduction
\item Axiomatics --- Propositional Logic 
\item Natural Deduction --- Mainly Propositional Logic
\item Natural Deduction --- First-Order Logic
\item Cool-down --- Summary \&\ Higher-Order Logic Introduction
\end{enumerate}

For the axiomatics we use a formalization in Isabelle of ``Hilbert Axioms'' based on David Hilbert's \emph{Die Grundlagen der Mathematik} 1928 and used in Alonzo Church's  \emph{Introduction to Mathematical Logic} 1956, cf.\ \cite{church1956introduction} page 163 (the system P1 also borrows from Gottlob Frege 1879, John von Neumann 1927 and in particular Mordchaj Wajsberg 1939).

 



It is a good exercise for the students to enter and verify the axioms in Isabelle:

\noindent
\includegraphics[trim=-25mm 12mm -25mm 22mm,clip,width=\textwidth,height=\textheight,keepaspectratio,page=2]{Hilbert}

\noindent
Most students find it straightforward to enter the formulas in Isabelle.
In the above Isabelle proof we have used the Isabelle's simplifier to prove the 9 axioms (\textsf{simp\_all}) but in general a proof method like Isabelle's classical tableau prover (\textsf{blast}) is required (for example, the simplifier does not succeed for $(A \rightarrow B \rightarrow \textit{False}) \rightarrow B \rightarrow A \rightarrow \textit{False}$).
Our reason for using the simplifier is that it is sufficient for the proofs to follow.

The following formalization of the axiomatics is based on ``Propositional Proof Systems'' by Julius Michaelis and Tobias Nipkow (Archive of Formal Proofs 2017) \cite{PropositionalProofSystemsAFP}.

\newpage

First the syntax and the semantics is defined and a small lemma is proved ($A \rightarrow A$):

\noindent
\includegraphics[trim=10mm 23mm 10mm 32mm,clip,width=\textwidth,height=\textheight,keepaspectratio,page=3]{Hilbert}

The datatype \textsf{form} defines the set of formulas.
The primitive recursive function \textsf{semantics} takes an interpretation and a formula.
We have chosen to use \textsf{if-then-else} expressions for the binary operators like in NaDeA.
We then define the provability predicate \textsf{OK} and easily prove soundness of the axiomatics (including the \emph{modus ponens} rule).

\noindent
\includegraphics[trim=10mm 30mm 10mm 0mm,clip,width=\textwidth,height=\textheight,keepaspectratio,page=4]{Hilbert}

It is again a good exercise for the students to enter the axioms, now for the provability predicate \textsf{OK} as shown above, just by looking at the normal syntax (and see how Isabelle highlights all mistakes in the process):

\noindent
\includegraphics[trim=-50mm 5mm -50mm 25mm,clip,width=\textwidth,height=\textheight,keepaspectratio,page=1]{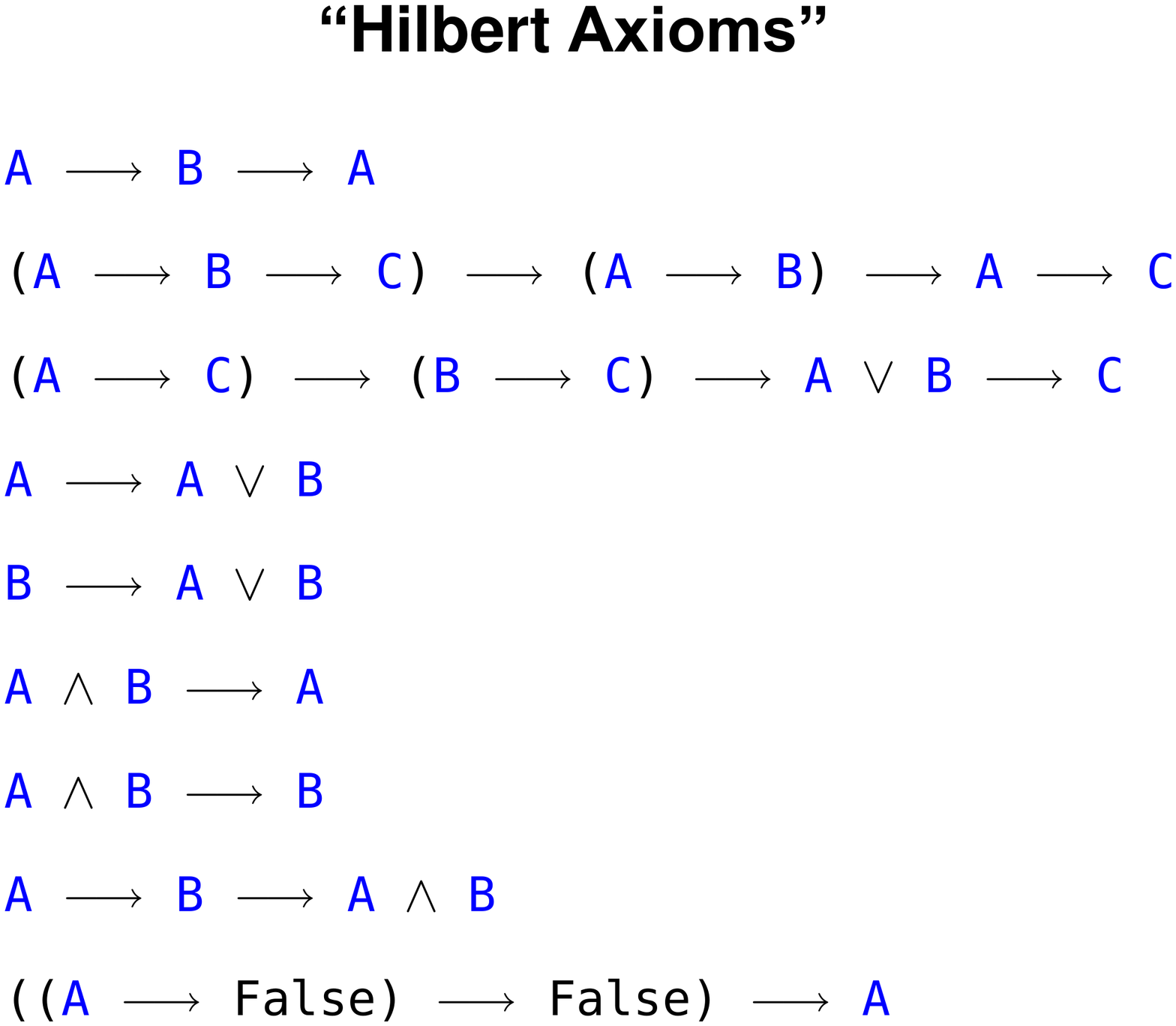}

We find that it is a good preparation to work in Isabelle with the much simpler axiomatics for propositional logic before considering the natural deduction proof system for first-order logic in NaDeA.

Of course it is not easy to carry out proofs in the axiomatics.
Even the proof of $A \rightarrow A$ requires 5 lines:
$$
\begin{array}{llr}
1. & \vdash (A \rightarrow (A \rightarrow A) \rightarrow A) \rightarrow (A \rightarrow A \rightarrow A) \rightarrow A \rightarrow A & \text{Axiom 2} \\[1ex]
2. & \vdash A \rightarrow (A \rightarrow A) \rightarrow A & \text{Axiom 1} \\[1ex]
3. & \vdash (A \rightarrow A \rightarrow A) \rightarrow A \rightarrow A & \text{MP 1, 2} \\[1ex]
4. & \vdash A \rightarrow A \rightarrow A & \text{Axiom 1} \\[1ex]
5. & \vdash A \rightarrow A & \text{MP 3, 4}
\end{array}
$$
This leads naturally to the notion of completeness:

\medskip

\noindent
\includegraphics[trim=-40mm 140mm -40mm 40mm,clip,width=\textwidth,height=\textheight,keepaspectratio,page=5]{Hilbert}

\medskip

A formal proof in Isabelle of the soundness and completeness theorems as given by \textsf{main} above is available for the interested students (about 1000 lines including other results).

We intend in the future also to let the students work on a formalization in Isabelle of a sequent calculus for propositional logic since this proof system can be taken as a simple automatic prover in contrast with the above axiomatics.

\newpage

\section{Student Evaluations of NaDeA}

The course evaluation spans all 13 weeks of the course, not just the couple of weeks spent on NaDeA.
We deem the overall evaluation relevant anyhow, and note that people tend to agree that the teaching material and course is good, and that they learn something from it.

\bigskip

Course evaluation results for 02156 Logical Systems and Logic Programming 2018 (Fall)
\begin{center}
\url{https://www.compute.dtu.dk/english/research/research-sections/algolog}
\end{center}

\bigskip

In total 17 out of 48 registered students answered the anonymous online form (35\%):

\bigskip

\noindent
\begin{tabulary}{\linewidth}{lLl}
1 & I think I am learning a lot in this course & 3.7 \\[4ex]
2 & I think the teaching method encourages my active participation & 3.5 \\[4ex]
3 & I think the teaching material is good & 3.6 \\[4ex]
4 & I think that throughout the course, the teacher has clearly communicated to me where I stand academically & 3.7 \\[4ex]
5 & I think the teacher creates good continuity between the different teaching activities & 3.7 \\[4ex]
6 & 5 points is equivalent to 9 hours/week --- I think my performance during the course is & 3.2 \\[4ex]
7 & I think the course description’s prerequisites are & 2.9 \\[4ex]
8 & In general, I think this is a good course & 3.7
\end{tabulary}%

\bigskip

The scale is ``strongly disagree'' (1) to ``strongly agree'' (5) except for 6 and 7 where it is ``much more'' (1) to ``much less'' (5) and ``too high'' (1) to ``too low'' (5), respectively.

\bigskip

To complement the course evaluation we asked students about feedback specifically on NaDeA.
We handed out to each of the 27 students present in classroom a paper sheet with the following text:

\bigskip

\hrulefill

\bigskip

Questionnaire --- Natural Deduction Assistant (NaDeA)

\

Advantages
\begin{itemize}
\item The proof system is formally proved sound and complete.
\item The structured environment makes one focus on the proof development process.
\item It forces one to input well-formed formulas and use applicable rules only.
\end{itemize}

Please add at least one new item:

\newpage

Disadvantages
\begin{itemize}
\item Mouse-clicking can be tedious.
\item It can be difficult to know whether the shortest proof has been achieved.
\item  For smaller proofs one can make more or less progress by clicking blindly and not understanding what is happening.
\end{itemize}

Please add at least one new item:

\bigskip

\hrulefill

\bigskip

12 students returned the paper sheet in box (some students informed us that they answered based on discussions in a small ad hoc group). 

One student wanted the possibility to change parts of a formula without having to undo all the way back to the step where it was introduced and  rebuilding the remaining proof.
This would require detecting how much of the proof is still correct, i.e. which applications of rules should be invalidated and determining what to do with those.
While cumbersome at times, the current requirement of undoing is a much simpler solution, both to implement and for students to understand.
One student appreciated that you can always undo to all previous states, which is contrary to many applications, even text editors, where performing an action after undoing means that the previous undos cannot be redone.

A related requested feature was the trimming of finished proofs to remove all undo steps, such that the proofs could be undone and redone without detours.
We have an external AWK script that does this, which is also useful to prepare proofs for presentation purposes, and integrating the feature into the system could make sense.

Another request was better scaling of the interface with the window width, and we miss this ourselves as well.
Especially for presentation purposes when the projector is narrower than the common laptop screen.
In the classroom where everyone uses laptops, it has not appeared to be a problem.

Finally someone requested a dark theme, that is, light-colored text on a dark background, while others praised the interface and especially the font.

The biggest gripe observed in the classroom was the need to click many times to input formulas.
As mentioned this problem was known to us already.
Another problem was confusion around the use of functions with no arguments as constants, but this problem is not specific to NaDeA.

Although NaDeA has a number of help/tutorial/exercises pages, we intend to provide more in-depth explanations in future versions.

\newpage

\section{Exercises in NaDeA}

In the classroom we ask the students to prove the following formulas (in addition to the ``standard''  example $A \rightarrow A$ used in the online tutorial):

\medskip

$$
\begin{array}{@{}ll}
\bot \rightarrow \bot
& \text{Test 1}
\\[2ex]
\bot \rightarrow A
& \text{Hint 1}
\\[2ex]
(A \rightarrow B) \rightarrow A \rightarrow B
& \text{Test 2}
\\[2ex]
A \rightarrow (A \rightarrow B) \rightarrow B
& \text{Hint 2}
\\[2ex]
A \land (A \rightarrow B) \rightarrow B
& \text{Test 3}
\\[2ex]
A(c) \land (A(c) \rightarrow \forall x A(x)) \rightarrow \forall x A(x)
& \text{Hint 3}
\\[2ex]
\forall x A(x) \rightarrow A(c)
& \text{Test 4}
\\[2ex]
A(c) \rightarrow \exists x A(x)
& \text{Hint 4}
\\[2ex]
A \rightarrow B \rightarrow A
& \text{Test 5}
\\[2ex]
(A \rightarrow B \rightarrow C) \rightarrow (A \rightarrow B) \rightarrow A \rightarrow C
& \text{Hint 5}
\\[2ex]
A \rightarrow (A \rightarrow \bot) \rightarrow \bot
& \text{Test 6}
\\[2ex]
((A \rightarrow \bot) \rightarrow \bot) \rightarrow A
& \text{Hint 6}
\\[2ex]
(A \land B) \rightarrow C \rightarrow (A \land C)
& \text{Test 7}
\\[2ex]
((A \rightarrow \bot) \lor (B \rightarrow \bot)) \rightarrow (A \land B) \rightarrow \bot
& \text{Hint 7}
\\[2ex]
\forall x \forall y A(x,y) \rightarrow \forall x A(x,x)
& \text{Test 8}
\\[2ex]
\forall x A(x) \rightarrow \exists x A(x)
& \text{Hint 8}
\\[2ex]
\exists x (A(x) \rightarrow \forall x A(x))
& \text{Test 9}
\\[2ex]
\forall x (\neg r(x) \rightarrow r(f(x))) \rightarrow \exists x (r(x) \land r(f(f(x))))
& \text{Hint 9 --- Optional.}
\end{array}
$$

\medskip

\noindent
Solutions are available online in NaDeA either as so-called hints in the Help window or as so-called tests in the Load window in NaDeA. A test resumes from the last proof state but a hint replays from the first proof state (one can click Undo to show the proof states).

However, we do not provide the solution to the final formula, in order to challenge the best students (we have a solution with 42 lines in NaDeA). The formula is discussed on page 128 of the Handbook of Tableau Methods (Kluwer Academic Publishers 1999):
\begin{quote}
\emph{If every person that is not rich has a rich father,
 then some rich person must have a rich grandfather.}
\end{quote}
The formalization uses $r$ for \emph{rich} and $f$ for \emph{father}.

\newpage

In the course the students must individually hand in 4 assignments with several quesions about logical systems and logic programming. With respect to NaDeA the students are asked to prove the following 5 formulas:

\begin{enumerate}
\item $A \land B \rightarrow B$
\item $A(c,c) \rightarrow \exists x \exists y A(x,y)$
\item $(\forall x A(x) \lor \forall x B(x)) \rightarrow \forall x (A(x) \lor B(x))$
\item $A \lor (A \rightarrow \bot)$
\item $(A \rightarrow B) \lor (B \rightarrow C)$
\end{enumerate}

Solutions to the assignments are not provided to the students. Instead we provide detailed individual feedback. Both the assignment with the NaDeA questions and the final written course exam are pending.

In case a student gets stuck with the last formula in the assignment,  $(A \rightarrow B) \lor (B \rightarrow C)$, some assistance can be provided as Example 1 and Example 2.

\subsection*{Example 1}

\includegraphics[scale=.42]{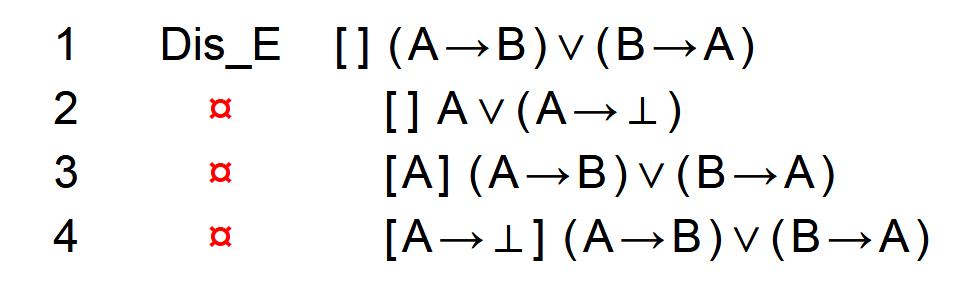}

\subsection*{Example 2}

\includegraphics[scale=.42]{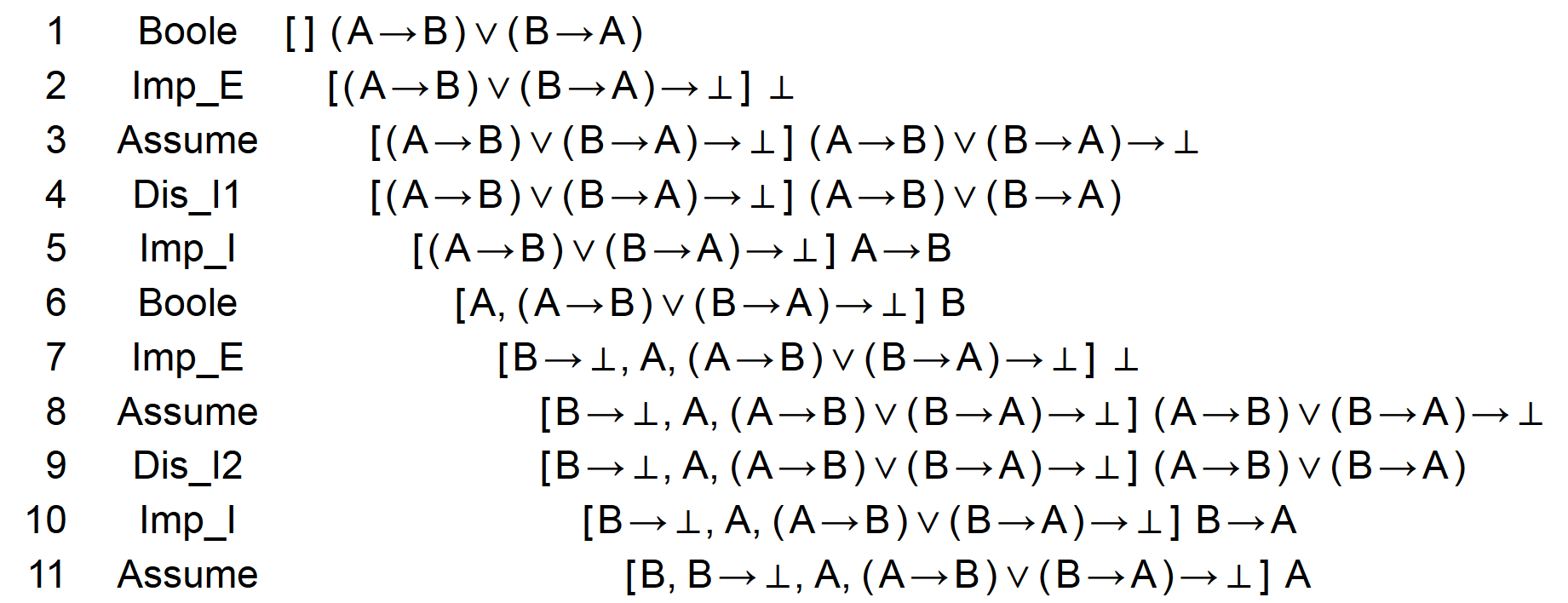}

\

\noindent
We note that although the solution for $(A \rightarrow B) \lor (B \rightarrow C)$ is similar to Example 1 as well as Example 2, at least a non-trivial observation is necessary in order to complete the proof in NaDeA.

\newpage

\section{Conclusion}

NaDeA has been used for teaching first-order logic to hundreds of computer science bachelor students.
We have discussed its advantages and disadvantages as a tool for teaching logic.
In general NaDeA has been a success in our bachelor course but we intend to provide more in-depth explanations in future versions of NaDeA.
However, in our experience it is more or less mandatory with a number of exercise sessions with skilled teaching assistants, say, one teaching assistant for a class of 40 bachelor students.
Advanced master or PhD students usually manage without help from teaching assistants.
For example, NaDeA has recently been used without a teaching assistant by a class of mainly PhD students at the 29th European Summer School in Logic, Language, and Information (ESSLLI), University of Toulouse, France, 17-28 July 2017:
\begin{center}
\url{https://www.irit.fr/esslli2017/courses/24.html}
\end{center}

\noindent
Proof assistants such as Isabelle allow for many kinds of reasoning that go beyond natural deduction
and their interfaces, of course, account for that.
In NaDeA, on the other hand, there are no distractions -- all buttons and texts in NaDeA have to do with natural deduction. The structured environment provided by NaDeA, based on clicking buttons instead of textual input, makes it possible for students to focus on the proof development process.
Since NaDeA only allows input of well-formed formulas and application of applicable rules, the student does not have to worry about neither syntax nor well-formedness errors possible for instance in Isabelle.
Conversely, experienced users may feel slightly inhibited by the system as textually inputting a formula is often faster than using the mouse.
Furthermore for very simple proofs, students may be able to find them by clicking blindly and without understanding what they are doing, since only the applicable rules are shown.
This is not a problem for larger proofs.

As future work we consider developing more teaching materials for NaDeA and making further evaluations of NaDeA as a tool for teaching logic.

\section*{Acknowledgements}

We thank Alexander Birch Jensen for collaboration on the initial development of NaDeA and we thank John Bruntse Larsen and Stefan Berghofer for fruitful discussions.
We also thank the anonymous reviewers for their comments. 

\bibliographystyle{eptcs}
\bibliography{references}

\newpage

\end{document}